\documentclass[doublespacing]{elsart}

\usepackage{graphicx}

\usepackage{amssymb}

\usepackage{amsmath}

\begin{document}

\begin{frontmatter}



\title{
A quantum critical superconducting phase transition in
quasi-two-dimensional systems with Dirac electrons
\thanksref{CNPq}
}
\thanks[CNPq]{This work has been
supported in part by CNPq and FAPERJ. ECM has been
partially supported by CNPq. LHCMN has been supported by CNPq.}


\author
{E. C. Marino},
\ead{marino@if.ufrj.br}
\author
{Lizardo H. C. M. Nunes}


\begin{abstract}
We present a theory describing the superconducting (SC) interaction
of Dirac electrons in a quasi-two-dimensional system consisting of a
stack of $N$ planes. The occurrence of a SC phase is investigated
both at $T=0$ and $T\neq 0$. At $T=0$,
we find a quantum phase transition connecting the normal
and SC phases. Our theory qualitatively
reproduces the SC phase
transition occurring in the underdoped regime of the high-Tc
cuprates. This fact points to the possible relevance of Dirac
electrons in the mechanism of high-Tc superconductivity.
\end{abstract}

\begin{keyword}
Dirac electrons
\sep superconductivity
\sep quantum criticality

\PACS
74.20-z
\sep 71.10.Hf
\end{keyword}

\end{frontmatter}

\section{Introduction}\label{int}

There are many condensed matter systems in one and two
spatial dimensions containing electrons that may be described by a
relativistic, Dirac-type lagrangian, namely Dirac electrons. Among these
we may list the high-Tc cuprates, graphene sheets and dichalchogenides \cite{geral}.
Even though these are evidently non-relativistic systems
these materials have special points in the
Brillouin zone where two bands touch in a single point around which
the electron dispersion relation behaves as $\epsilon(\vec
k) = v_{ \rm F} |\vec k|$. The elementary excitations around such a point
are Dirac electrons. They are, after all, a result of the
electron-lattice interaction.

We present here, a theory describing the superconducting
interaction of Dirac electrons associated to two distinct Dirac
points \cite{ml}.
We show that, at $T=0$, the system presents a quantum critical point
separating the normal and superconducting phases and determine the
superconducting gap as a function of the coupling constant.
The quantum phase transition occurring in our model and the behavior
of $T_c$ around the quantum critical point qualitatively reproduce
very well the superconducting transition in the high-Tc cuprates in
the underdoped region. This suggests that Dirac electrons may play
an important role in the mechanism of high-Tc superconductivity.

We consider a quasi-two-dimensional electronic system consisting of
a stack of planes containing two Dirac points. In addition, we
introduce an internal index $a=1,...,N$, supposed to characterize
the different planes to which the electrons may belong. The electron
creation operator, therefore, is given by $\psi^\dag_{i\sigma a}$,
where $i=1,2$ are the Dirac indices, corresponding to the two Fermi
points, $\sigma = \uparrow,\downarrow$, specifies the z-component of
the electron spin and $a=1,...,N$ labels the electron plane.
The complete lagrangian we will consider is given by
$$
\mathcal{ L }  ={\rm i} \overline\psi_{ \sigma a} \not\! \partial \
\psi_{ \sigma a} + \frac{\lambda}{N} \left (\psi^\dag_{1\uparrow a} \
\psi^\dag_{2\downarrow a} + \psi^\dag_{2\uparrow a} \
\psi^\dag_{1\downarrow a}  \right )
$$
\begin{equation}
  \times \left  (\psi_{2\downarrow b} \
\psi_{1\uparrow b} + \psi_{1\downarrow b} \ \psi_{2\uparrow b}
\right ), \label{L}
\end{equation}
where $\lambda > 0$ is a constant that may depend on some external control
parameter, such as the pressure or the concentration of some dopant.

We now introduce a Hubbard-Stratonovitch complex scalar field
$\sigma$, in terms of which the lagrangian becomes

$$
\mathcal{ L } \left[ \Psi, \sigma \right]
 =
{\rm i} \overline\psi_{ \sigma a} \not\! \partial \ \psi_{ \sigma a}
- \frac{N}{ \lambda }\ \sigma^{ * }  \sigma
$$
\begin{equation}
- \sigma^{ * } \left( \psi_{2\downarrow b} \ \psi_{1\uparrow b}
+ \psi_{1\downarrow b} \ \psi_{2\uparrow b}\right)
- \sigma \left(\psi^\dag_{1\uparrow a} \
\psi^\dag_{2\downarrow a} + \psi^\dag_{2\uparrow a} \
\psi^\dag_{1\downarrow a} \right).
\label{lsigpsi}
\end{equation}
>From this we obtain the field equation for the auxiliary field:
$\sigma = - \frac{\lambda}{N}\ \left (\psi_{2\downarrow a} \ \psi_{1\uparrow a} +
\psi_{1\downarrow a}\ \psi_{2\uparrow a} \right )
$ The vacuum expectation value of $\sigma$ is an order parameter for the superconducting phase.

Integrating on the fermion fields , we obtain the effective action
\begin{equation}
S_{eff} \left[ \sigma \right]  = \int d^{ 3 } x \left( -
\frac{N}{\lambda} |\sigma |^2
 \right) - {\rm i}2 N \rm{Tr} \ln \left [
1 + \frac{|\sigma |^2}{\Box}\right] \label{seff}
\end{equation}
Let us consider firstly $T=0$. In this case,
we get the renormalized
effective potential per plane corresponding to (\ref{seff}):
\begin{equation}
V_{ { \rm eff},R } \left( |\sigma| \right) = \frac{ |\sigma|^2 }{
\lambda_R } - \frac{ 3 \sigma_0 }{ 2 \alpha } |\sigma|^2 + \frac{
2}{ 3 \alpha } |\sigma|^3 , \label{veffr}
\end{equation}
where $\lambda_R$ is the (physical) renormalized coupling
and $\sigma_0$ is an arbitrary finite scale, the renormalization
point.

Studying the minima of the previous expression,
we can infer that the ground state of the system will be
\begin{equation}
 \Delta_0 =
 \left \{  \begin{array}{c}
 0 \ \ \ \ \ \ \ \ \ \ \ \ \
 \lambda_R < \lambda_c \\    \\
 \alpha\left(\frac{1}{\lambda_c} - \frac{  1  }{ \lambda_R
}\right)
  \ \ \ \ \ \ \ \ \  \lambda_R > \lambda_c
       \end{array} \right .
 \label{s00},
\end{equation}
where $\Delta = |\sigma|$.
Expression (\ref{s00}) implies that the system undergoes a
continuous quantum phase transition at the quantum critical point
$\lambda_c =  4 \pi v^2_{ \rm{F} } / 3 \sigma_0 $.
separating a normal from a superconducting phase.

We turn now to finite temperature effects. Using a large $N$ expansion
and evaluating (\ref{seff}) at $T\neq 0$, we find the effective potential,
whose minima provide a general expression for the
superconducting gap as a function of the temperature, namely
\begin{equation}
\Delta (T) = 2 T \cosh^{- 1}\left[ \frac{e^{\frac{\Delta_0}{2T}}}{2}
\right ], \label{gap1}
\end{equation}
where $\Delta_0$ is given by (\ref{s00}). From
(\ref{gap1}) we can verify that indeed $\Delta (T=0)=\Delta_0$. Also
from the above equation, we may determine the critical temperature
$T_c$ for which the superconducting gap vanishes. Using the fact
that $\Delta (T_c)= 0$, we readily find from (\ref{gap1})
\begin{equation}
T_c =  \frac {\Delta_0}{2 \ln 2} \label{gap3}.
\end{equation}

\begin{figure}[ht]
\centerline {
\includegraphics
[clip,width=0.9\textwidth
,angle=-90
] {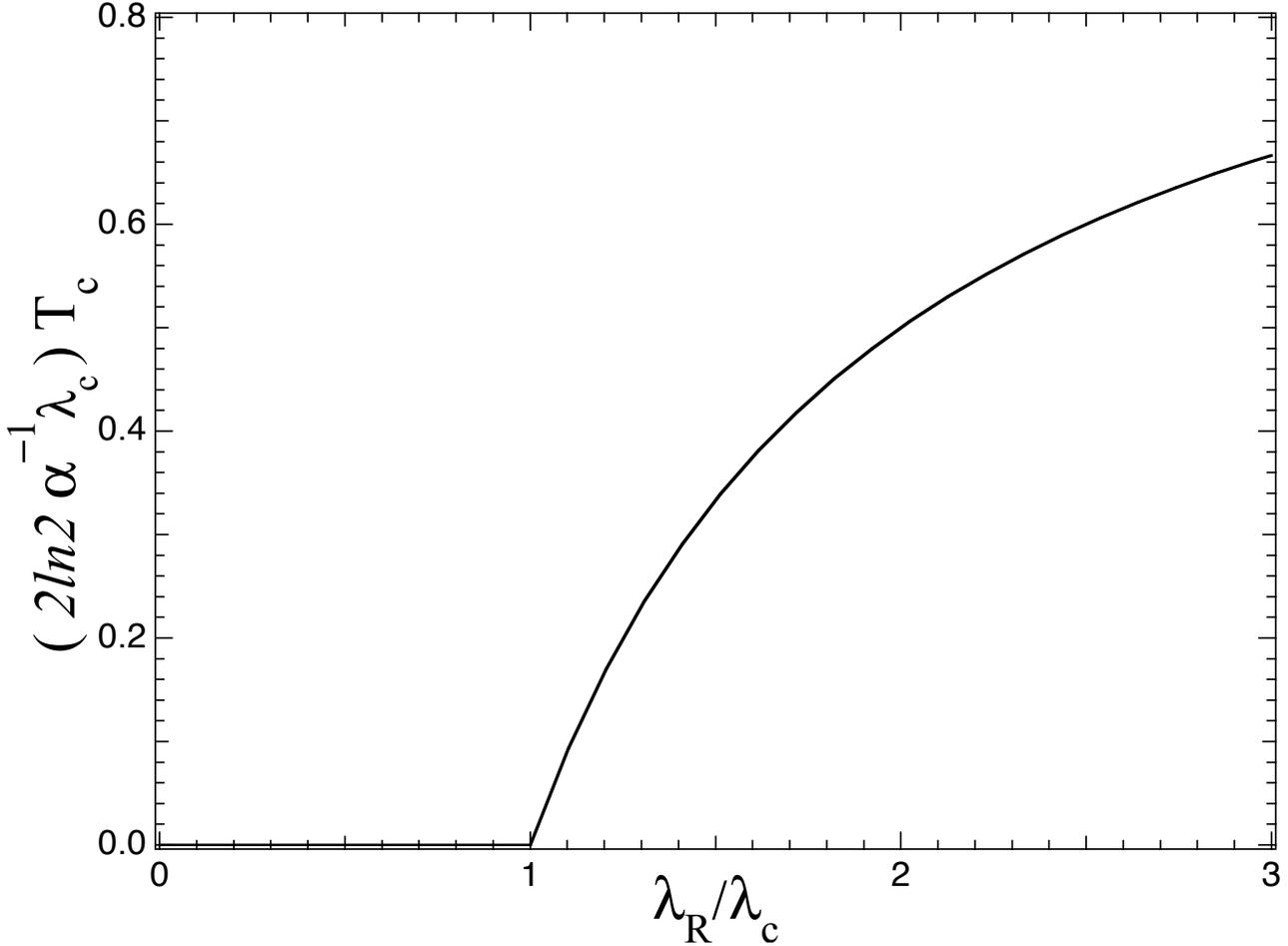} }
\caption{The superconducting critical
temperature $ T_c $ as a function of the renormalized coupling  $
\lambda_R  $.} \label{fig1}
\end{figure}

In Fig. \ref{fig1}, using (\ref{s00}) and (\ref{gap3}), we display
$T_c$ as a function of the coupling constant. This qualitatively
reproduces the superconducting phase transition of the high-Tc
cuprates in the underdoped region. Since our theory describes the
generic superconducting interaction of two-dimensional Dirac
electrons, we may see this result as an indication of the possible
relevance of this type of electrons in the high-Tc mechanism.

 In terms of the critical temperature, we may also express the gap as
\begin{equation}
\Delta (T) = 2 T \cosh^{- 1}\left[
2^{\left(\frac{T_c}{T}-1\right)}\right ] \label{gap4} \ .
\end{equation}
Near $T_c$, this yields
\begin{equation}
\Delta (T) \stackrel{T\lesssim T_c}{\sim} 2 \sqrt{2\ln 2}\ T_c\left
(1- \frac{T}{T_c} \right)^{\frac{1}{2}}\label{gap5} \ ,
\end{equation}
which presents the typical mean field critical exponent $1/2$.

Finally, we would like to make two remarks. Firstly, both the gap $\Delta(T)$
(and hence the critical temperature) and
the renormalized effective potential do not depend on the arbitrary
renormalization point $\sigma_0$. This can be seen by a renormalization group analysis
\cite{ml}. The theory does not predict the value of $\lambda_c$,
it has to be determined experimentally. Second, we can show that the results, obtained in mean field,
are robust against quantum fluctuations \cite{ml}.






\end{document}